\documentclass[12pt]{iopart}
\usepackage{iopams}  
\usepackage[T1]{fontenc}
\usepackage{array}
\usepackage{braket}
\usepackage{booktabs}
\usepackage{graphicx}
\usepackage{subfig}
\usepackage[linesnumbered,ruled,vlined]{algorithm2e}
\usepackage{color}
\usepackage{ulem}
\usepackage{url}
\usepackage{textcomp}
\usepackage{bbold}
\usepackage{citesort}
\newcommand{\nket}[1]{\left|#1\right)}
\newcommand{\qs}{$\mathcal{Q}$System}

\begin{document}
\title{$\mathcal{Q}$System: bitwise representation for quantum circuit simulations}

\author{E. C. R. Rosa$^1$ and B. G. Taketani$^2$}

\address{$^1$ Departamento de Inform\'atica e Estat\'istica - PPGCC, Universidade Federal de Santa Catarina, Florian\'opolis, 88040-900, Brazil.}
\address{$^2$ Departamento de F\'isica, Universidade Federal de Santa Catarina, Florian\'opolis, 88040-900, Brazil.}

\ead{\mailto{evandro.crr@posgrad.ufsc.br}}

\begin{abstract}
We present $\mathcal{Q}$System, an open-source platform for the simulation of
quantum circuits focused on bitwise operations on a Hashmap data
structure storing quantum states and gates. $\mathcal{Q}$System is
implemented in C++ and delivered as a Python module, taking advantage of the
C++ performance and the Python dynamism. The simulator’s API is designed to
be simple and intuitive, thus streamlining the simulation of a quantum
circuit in Python. The current release has three distinct ways to represent
the quantum state: vector, matrix, and the proposed bitwise. The latter
constitutes our main results and is a new way to store and manipulate both
states and operations which shows an exponential advantage with the amount of
superposition in the system's state. We benchmark the bitwise representation
against other simulators, namely Qiskit, Forest SDK QVM, and Cirq.
\end{abstract}


\section{Introduction}
Quantum computation is a new paradigm of computing which uses quantum
phenomena, such as superposition and entanglement, to efficiently tackle
classes of problems out of reach for classical computers. Since the seminal
works of Shor~\cite{shor:1997} and Grover~\cite{Grover1996}
showcasing relevant quantum algorithms with advantage over their classical
counterparts, the field has seen significant efforts both in theory and
experiments \cite{Farhi:2014,Peruzzo:2014,DallaireDemers:2016}. The search
for other possible algorithms is a very active field as well their
implementations or simulations
\cite{Biamonte:2017,Govia:2017,Taketani:2018,Hempel:2018,Schmit:2020}.

In this scenario, the simulation of quantum dynamics and small quantum
circuits in a classical computer is of central importance. These may be used
to validate computing models \cite{AspuruGuzik:2005,Sieberer:2019}, simulate
physical theories
\cite{Moll:2018,Gustafson:2019,Messinger:2019,Lamm:2019,Wang:2019} and even
to define the break-even point, when quantum computers match what is
classically achievable \cite{Preskill:2018}. To this end, several quantum
circuit simulators have been developed in the past years, both from academic
and industrial players
\cite{Johansson2012,Johansson:2013,smith_practical_2016,Gidney2018,Steiger2018,qiskit,Jones:2019,Killoran2019}.
In common to these approaches is the drive to improve the memory and time
cost of such simulations. Owing to the predicted exponential scaling of these
resources with the number of qubits, efficiency is at the core of previous
implementations, which try to push the boundary of what can be simulated
\cite{Pednault:2017,Arute2019,Pednault:2019}.

Here we introduce a new approach to quantum circuit simulations that can
display exponential advantage over other simulators. Our approach is based on
Hashmaps to store quantum states and gates, and uses bitwise 
operations to implement gates. For quantum states that can be well described
by a small, yet unknown set of basis states, we argue that the small number of keys
in their respective Hashmap leads to an exponential speedup. This is further
observed in benchmark tests against a number of popular simulators. We
implement this approach on $\mathcal{Q}$System, a quantum circuit simulator
which was developed to support the study and construction of quantum
algorithms, quantum protocols, error correction codes and other quantum
applications that can be described as a quantum circuit. The
$\mathcal{Q}$System's API is developed as a Python module, wrapping a C++
code, with methods resembling common quantum computing
language/terminology, making it easy the transcribe any quantum circuit.
Besides the bitwise representation, $\mathcal{Q}$System also provides a
vector representation of pure quantum states and a density matrix
representation for simulations of circuits with quantum errors, including
predefined methods for quantum error correction.

The remainder of this paper is organized as follows: In \Sref{sec:qc}, a brief introduction to quantum computation is given. \Sref{sec:qys} presents the $\mathcal{Q}$System simulator and the bitwise representation is described in \Sref{sec:bw}. The benchmarks are discussed in \Sref{sec:benchmark} and our concluding remarks are given in \Sref{sec:conclusion}.

\section{Quantum computing}
\label{sec:qc}
The quantum computing basic operating unit is the quantum bit, or qubit. Like the classical bit, the qubit can be at the state 0, represented by $\ket{0}$, and 1, represented by $\ket{1}$, which form a basis for the qubit description and manipulation. However, unlike the classical bit, a qubit can also be described as a convex sum (or superposition) of its basis states represented as $\alpha\ket{0}+\beta\ket{1}$, with $\alpha$ and $\beta \in \mathbb{C}$ and $|\alpha|^2+|\beta|^2 = 1$~ \cite{nielsen_quantum_2010}. Measuring the qubit in this computational basis destroys this superposition and collapses the quantum state to either $\ket 0$ or $\ket 1$ with probabilities $|\alpha|^2$ and $|\beta|^2$, respectively. Therefore a $2$-dimensional normalized complex vector can be used to represent the state of a qubit. This is often the path used for numerical simulations of quantum mechanics.

For an $n$-qubit system, the above ideas can be generalized and a unit vector in $\mathbb{C}^{2^n}$ is used to represent the quantum state. It is immediately clear that the dimension of the vector space scales exponentially with system size, rendering simulations of systems larger than a few dozen qubits intractable. 

The computational basis is defined by a set of basis states where each qubit is in either the $\ket 0$ or $\ket 1$ states (no superpositions allowed in this basis states). For a $3$-dimensional system these would be
\begin{eqnarray}
\matrix{
\ket0 = \nket {000}, & \ket1=\nket{001}, & \ket2=\nket{010}, \cr
\ket3 = \nket {011}, & \ket4=\nket{100}, & \ket5=\nket{101}, \cr
\ket6 = \nket {110}, & \ket7=\nket{111}, 
}
\end{eqnarray}
where we have introduced the notation $\ket n$ as a useful representation of
the computational basis states in terms of integer numbers and $\nket \cdot$
to represent these in terms of individual qubit
states. An arbitrary system state, $\ket{\psi}=\sum_{i=0}^{2^n-1}
\alpha_i\ket{i}$, can the be represented as a $2^n-1$ dimensional unit
vector with $i$-component given by $\alpha_i$. In $\mathcal{Q}$System this
will be called a vector representation.

For systems with classical error, the quantum state is described by the so-called density matrix. If the system has a classical probability $p_i$ of being found in a given quantum state $\ket{\psi_i}$, its density matrix is $\rho = \sum_{i}p_i\ket{\psi_i}\mkern-7mu\bra{\psi_i}$. In a numerical implementation this can be mapped to a matrix representation where $\ket{\psi}\mkern-7mu\bra{\phi}$ is the outer product of a column vector $\ket\psi$ and a row vector $\ket\phi$. Such description allows one to include errors in the computation and study, \textit{e.g.}, quantum error correction schemes~\cite{nielsen_quantum_2010}. In $\mathcal{Q}$System this will be called a matrix representation.

Operations on the system state can take three forms, quantum gates, quantum channels and measurements. Quantum gates are unitary operations acting on a subset of the qubits. These can be represented by $2^n\times2^n$ matrices. As an example, the Hadamard gate acting on states $\ket{0}$ and $\ket{1}$ of a single qubit system is given by
\begin{eqnarray}
    H\ket{0} &= 
    \frac{1}{\sqrt{2}}\left[ \matrix{ 1 & 1 \cr 1 & -1 } \right] \left[ \matrix{ 1 \cr 0 } \right] =
    \frac{1}{\sqrt{2}}\left[ \matrix{1 \cr 1} \right] &= 
    \frac{1}{\sqrt{2}}(\ket{0}+\ket{1}) \equiv \ket{+} \label{eq:h0} \\
    H\ket{1} &= 
    \frac{1}{\sqrt{2}} \left[ \matrix{ 1 & 1 \cr 1 & -1 } \right] \left[ \matrix {0 \cr 1 } \right] = 
    \frac{1}{\sqrt{2}} \left[ \matrix{ 1 \cr -1} \right] &=
    \frac{1}{\sqrt{2}}(\ket{0}-\ket{1}) \equiv \ket{-} \label{eq:h1}
\end{eqnarray}

In quantum computation, the most widely used measurement type is the projective measurement on the eigenbasis of a given observable, $\mathcal{O}$, and the measurement statistics determines the outcome of a given computation. This measurement projects the state, $\rho$, into one of the basis states with a probability given by $\Tr(\rho \mathcal{O})$.

The last operation that can be performed is a quantum channel. These are used to account for errors in the computation and lead to classical uncertainty in the final state and can be described by $\rho^\prime=\sum_i K_i\rho K_i^\dagger$, where $K_i$ are the Kraus operators satisfying $\sum_i K_i^\dagger K_i=\mathbb{1}$ for trace-preserving channels. This describes non-unitary evolutions and is the standard way to study error propagation and quantum error correction. We refer the interested reader to References \cite{gottesman_stabilizer_1997,nielsen_quantum_2010} for a detailed discussion.

Naturally, there is a large set of more specific tools, protocols and algorithms in quantum computation, many of which are implemented in \qs . In the following we will introduce the \qs\, simulator, focusing on the bitwise representation and on a few of its methods. The complete list of methods is available in Reference~\cite{Rosa2019}.

\section{The Simulator}
\label{sec:qys}
\qs\, is a quantum circuit simulator for Python designed with three central
concepts: (i) efficiency, (ii) simple methods and (iii) general use. Altough
the simulator is delivered as a Python module, it was developed in C++ for
improved efficiency. The software SWIG~\cite{beazley_swig:_1996} was used to
wrap the C++ code and generate an interface with the Python interpreter. This
is done by a configuration file and no modification is needed in the C++
source code. With this, the core C++ code is independent of the Python
library and can be used standalone with few modifications. Our main result,
the bitwise representation is a powerful tool to improve efficiency in some
cases, and will be discussed in \Sref{sec:bw}.

To design a toolbox with methods that are simple and intuitive to call, we
tried to keep the coding as close as possible to the natural use of the
English language when referring to a given tool. As an example, applying a
controlled phase gate of $\theta$ to target qubit $t$, with control qubits
$c_1$ and $c_2$ is done with the function \texttt{cphase}$\left(\theta, t,
\left[c_1,c_2\right]\right)$.

Typical quantum computing core operations have been implemented in \qs, as
well as more sophisticated functions, such as quantum Fourier
transform~\cite{Coppersmith:2002} and some quantum error correction
protocols~\cite{gottesman_stabilizer_1997}. The simulator can thus be used
for a number of applications, including quantum algorithms, quantum protocols
and error correcting codes. The evolution of a state is described in two
classes: (i) the QSystem class, which names the simulator, holds and controls
the quantum state, applying quantum gates and performing measurements; and
(ii) the Gate class, used to create quantum gates different from the default
set of gates in the QSystem class.

\subsubsection*{Representation.}
The $\mathcal{Q}$System simulator has three forms to represent the quantum
state, the bitwise and vector for pure states and matrix for simulations of
noisy systems. The bitwise representation is detailed in
Section~\ref{sec:bw}. As the vector and matrix representation are a straight
forward implementation of state vectors and density matrices using sparse
matrices (see \Sref{sec:qc}), we will not describe these representations
here. For linear algebra operations, such as matrix multiplication and tensor
product, we used the template-based C++ Armadillo
library~\cite{sanderson_armadillo:_2016,sanderson_practical_2019}.

\subsubsection*{Quantum gates.}
The QSystem class has several methods to apply quantum gates. Single-qubit
gates are summarized in \Tref{tab:evol} and Multiple-qubits gates in
\Tref{tab:cgate}. Other predefined methods include \texttt{qft()}, used to
apply a Quantum Fourier Transformation, and \texttt{apply()} used to apply a
user-defined quantum gate, created by the Gate class.
\begin{table}[hb]
  \caption{%
  Single-qubit gates predefined in QSystem. The last column refers to QSystem's method and the required parameters for implementations.
  }
  \label{tab:evol}
  \centering
  \begin{tabular}{lcl}
    \toprule
    Gate name  & Matrix form & Method \\
    \midrule
      \vspace{.5mm}
    Pauli X    & $\left[ \matrix{0 & 1 \cr 1 & 0 }\right]$                                                                  & \texttt{evol('X', qubitID)} \\
      \vspace{.5mm}
    Pauli Y    & $\left[ \matrix{0 & -i \cr i & 0 }\right]$                                                                 & \texttt{evol('Y', qubitID)} \\
      \vspace{.5mm}
    Pauli Z    & $\left[ \matrix{1 & 0 \cr 0 & -1 }\right]$                                                                 & \texttt{evol('Z', qubitID)} \\
      \vspace{.5mm}
    Hadamard   & $\frac{1}{\sqrt{2}}\left[ \matrix{1 & 1 \cr 1 & -1 }\right]$                                                                           & \texttt{evol('H', qubitID)} \\
      \vspace{.5mm}
    S Gate     & $\left[ \matrix{1 & 0 \cr 0 & i }\right]$                                                                  & \texttt{evol('S', qubitID)} \\
      \vspace{.5mm}
    T Gate     & $\left[ \matrix{1 & 0 \cr 0 & e^\frac{i\pi}{4} }\right]$                                                                                 & \texttt{evol('T', qubitID)} \\
      \vspace{.5mm}
    X rotation & $\left[\matrix{\cos\frac{\theta}{2} & -i\sin\frac{\theta}{2} \cr -i\sin\frac{\theta}{2} & \cos\frac{\theta}{2}}\right]$                 & \texttt{rot('X', $\theta$, qubitID)}  \\
      \vspace{.5mm}
    Y rotation & $\left[\matrix{ \cos\frac{\theta}{2} & -\sin\frac{\theta}{2} \\ \sin\frac{\theta}{2} & \cos\frac{\theta}{2}} \right]$                    & \texttt{rot('Y', $\theta$, qubitID)}  \\
      \vspace{.5mm}
    Z rotation & $\left[\matrix{ -e^{i\theta/2} & 0 \\ 0 & e^{i\theta/2} }\right]$                                                                        & \texttt{rot('Z', $\theta$, qubitID)}  \\
      \vspace{.5mm}
    U3 gate    & $\left[\matrix{ \cos\frac{\theta}{2} & -e^{i\lambda}\sin\frac{\theta}{2} \cr e^{i\phi}\sin\frac{\theta}{2} & e^{i(\lambda+\phi)}\cos\frac{\theta}{2}}\right]$ & \texttt{u3($\theta$, $\phi$, $\lambda$, qubitID)}       \\
      \vspace{.5mm}
    U2 gate    & $\frac{1}{\sqrt{2}} \left[\matrix{ 1 & -e^{i\lambda} \\ e^{i\phi} & e^{i(\lambda+\phi)} }\right]$                                                             & \texttt{u2($\phi$, $\lambda$, qubitID)}       \\
      \vspace{.5mm}
    U1 gate    & $\left[\matrix{ 1 & 0 \\ 0 & e^{i\lambda}}\right]$                                                                                & \texttt{u1($\lambda$, qubitID)}       \\
    \bottomrule
  \end{tabular}
\end{table}

\newcommand{\vcenteredinclude}[1]{\begingroup%
  \setbox0=\hbox{\includegraphics{#1}}%
  \parbox{\wd0}{\box0}\endgroup}

\begin{table}[hb]
  \caption{%
  Multiple-qubits gates predefined in QSystem. The last column refers to QSystem's method used to apply it. 
  }
  \label{tab:cgate}
  \centering
  \begin{tabular}{lcl}
    \toprule
    Gate name & Circuit form                    & Method \\
    \midrule
    \vspace{3pt}
    CNOT      & \vcenteredinclude{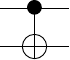} & \texttt{cnot(targetQubitID, [ctrlQubitID, $\dots$])}   \\
    \vspace{3pt}
    CPhase    & \vcenteredinclude{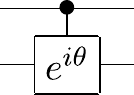}  & \texttt{cphase($e^{\theta}$, targetQubitID, [ctrlQubitID, $\dots$])} \\
    \vspace{3pt}
    SWAP      & \vcenteredinclude{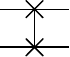} & \texttt{swap(qubitIDa, qubitIDb)}   \\
    \bottomrule
  \end{tabular}
\end{table}

The Gate Class has four static methods to create a custom gate: 
\begin{itemize}
\item \texttt{Gate.from\_matrix([$U_{00}$, $U_{01}$, $U_{10}$, $U_{11}$])}, used to create a single-qubit gate from an input matrix.
\item \texttt{Gate.from\_sp\_matrix(size, row, column, value)}, used to create an $n$-qubit gate from a sparse matrix. The sparse matrix $U$ is split into three parts, row, column, and value, where
\begin{equation}
U(\texttt{row}\texttt{[i]}, \texttt{column}\texttt{[i]}) = \texttt{value}\texttt{[i]}.
\end{equation}
\item \texttt{Gate.from\_func(func, $n$)}, used to create an $n$-qubit gate from a Python function. Considering a Python function \texttt{func()} that has a single unsigned integer as input and output, the matrix that represents the created gate is defined as
\begin{equation}
  \sum_{i=0}^{2^n-1} \ket{\texttt{func(}i\texttt{)}}\mkern-7mu\bra{i}.
\end{equation} 
\item \texttt{Gate.cxz\_gate(stabilizer, [ctrlQubitID, $\dots$])}, used to create a syndrome measurement gate, used in quantum stabilizer codes~\cite{gottesman_stabilizer_1997}.
\end{itemize}

For vector and matrix representations we use late evaluation to reduce the
time taken on matrix multiplications. When a gate is applied, the request is
registered in a list where each item describes the gates acting on each
qubit. When a new computation depends on any of its pending results, these
gates are applied generating the matrix for the unitary transformation.

\subsubsection*{Measurement.}

The simulator is able to perform measurements in the computational basis
(equivalent to a Pauli Z measurement) using the methods \texttt{measure(qubitID)}
and \texttt{measure\_all()}. Measurements on other basis can be achieved by
pre-measurement rotations, \textit{e.g.} a Pauli X measurement is performed
by applying a Hadamard gate before the measurement.

When a qubit is measured it collapses on the computational base state
corresponding to the measurement result. The qubit can be subsequently used
and measured again. Measurement results are stored in a list accessible via
the method \texttt{bits()}. Each element of this list holds the last
measurement outcome of each qubit, where the possible values are \texttt{0},
\texttt{1}, or \texttt{None} for unmeasured qubits.

\subsubsection*{Quantum errors.}
Quantum errors can be simulated within the matrix representation. Predefined error channels are:
\begin{itemize}
\item Bit flip, \texttt{flip('X', $\rho$, $p$)}: $\mathcal{E}(\rho|p) = (1-p)\rho + p(X\rho X)$.
\item Phase flip, \texttt{flip('Z', $\rho$, $p$)}: $\mathcal{E}(\rho|p) =(1-p)\rho + p(Z\rho Z)$.
\item Bit-phase flip, \texttt{flip('Y', $\rho$, $p$)}: $\mathcal{E}(\rho|p) =(1-p)\rho + p(Y\rho Y)$.
\item Amplitude damping, \texttt{amp\_damping($\rho$, $p$)}: 
\\$\mathcal{E}(\rho|p) = \left[\matrix{ 1 & 0 \cr 0 & \sqrt{1-p} }\right] \rho \left[\matrix{ 1 & 0 \cr 0 & \sqrt{1-p} }\right] + \left[\matrix{ 0 & 0 \cr \sqrt{p} & 0}\right] \rho \left[\matrix{ 0 & \sqrt{p} \cr 0 & 0 }\right]$.
\item Depolarizing channel, \texttt{dpl\_channel($\rho$, $p$)}: 
\\$\mathcal{E}(\rho|p) = \left(1-\frac{3p}{4}\right)\rho +\frac{p}{4}(X\rho X+Y\rho Y+Z\rho Z)$.
\end{itemize}
Where $p$ is the probability that an error occurs.

\section{The bitwise representation}
\label{sec:bw}

\subsubsection*{Bitwise operations.}
\label{sec:alg}
The bitwise operations underlying the bitwise representation presented here
are logical and shift operations on a set of individual bits of a bit string.
These treat a given number in its binary representation and operate on
it bit-by-bit, and will be used, \textit{e.g.}, to transform one basis state
to another. Logical bitwise operations used here include AND (``$\wedge$''),
OR (``$|$''), and XOR (``$\oplus$''). To illustrate
these, consider the following operation $\ket3$ OR $\ket 5$ in a 4-qubit
simulator. This gives
\begin{eqnarray}
\matrix{
&\ket3\quad( \nket{0011})\cr
\mathrm{OR}&\ket5\quad(\nket{0101})\cr
\hline\cr
=&\ket7\quad(\nket{0111})
}.
\end{eqnarray}

The AND (``$\wedge$'') is used to apply a selection mask to the bit string. This
mask is usually given by numbers of the form 
\begin{equation}
 2^n\equiv 01\overbrace{0\dots0}^n, 
\end{equation}
 with the associated bit string consisting of a bit 1 at position $n+1$ (from
 right to left) and all other bits 0. The mask can also be given by numbers of the form
\begin{equation}
 2^n-1\equiv 0\overbrace{1\dots1}^n, 
\end{equation}
 with a bit string consisting
of bits 1 for positions to the right of $n$ and all other bits set to 0.
For example, if we what to select the first three qubits of $\ket{22}$, 
we use the operation 
$\ket{22} \wedge (2^3-1)$, that gives
\begin{eqnarray}
  \label{eq:and}
\matrix{
&\ket{22} &(\nket{10110})\cr
\mathrm{AND}&2^3-1  &(00111)\cr
\hline\cr
=&\ket6 & (\nket{00110})
}.
\end{eqnarray}
$\mathcal{Q}$System indexes qubits from left to right starting at 0.

The XOR (``$\oplus$'') is used flip a bit. This way, if we what 
to apply a Pauli X gate on the $n$-th qubit of the estate $\ket{i}$ we use
the operation $\ket{i} \oplus 2^n \equiv X_n\ket{i}$, \textit{e.g.}, 
$\ket{10} \oplus 2^2$ gives
\begin{eqnarray}
  \label{eq:xor}
\matrix{
&\ket{9} &(\nket{1001})\cr
\mathrm{XOR}&2^2  &(0100)\cr
\hline\cr
=&\ket{14} & (\nket{1101})
}.
\end{eqnarray}

Shift operations move the entire bit string sideways a given number of times
and can be Shift left (``$\ll$'') and Shift right (``$\gg$''). Apply a Shift
left adds zeros to the right side of the bit string, multiplying the number
by $2^n$, \textit{e.g.}, $7 \ll 2 = 28$. It is useful to create masks,
\textit{e.g.}, $(1 \ll n) -1 = 2^n-1$ creates a mask to select the first $n$
bits, and $1 \ll n = 2^n$ can be used to flip the $n$-th bit. We can rewrite
\Eref{eq:and} as $\ket{22} \wedge ((1 \ll 3)-1) = \ket6$ and \Eref{eq:xor} as
$\ket{10} \oplus (1 \ll 2) = \ket{14}$. The Shift right consumes the first
bits, dividing (entire division) it by $2^n$, \textit{e.g.}, $7 \gg 2 = 1$.

\subsubsection*{State and gate representations.}
The bitwise representation uses Hashmaps to represent pure states and quantum gates~\cite{Cormen:2009}. While Hashmaps require heavy memory use, the concise way to represent sparse states and gates allows them to have significant gain in time, as will be seen in \Sref{sec:benchmark}. Pure states will be represented by the Hashmap \texttt{bwQubits}, which takes unsigned integers as \texttt{key} and maps it to a complex number. The \texttt{key} represents a base vector $\ket{i}$ of the system and the complex number its amplitude $\alpha_i$, with the equivalence
\begin{equation}
  \ket{\psi} = \sum_{i=0}^{2^n-1} \alpha_i\ket{i} \equiv \sum_{\textrm{key}\,\in\,\texttt{bwQubits}} \texttt{bwQubits}[ \textrm{key} ]\ket{\textrm{key}}.
\end{equation}

As with the sparse matrix representation, this representation only stores the
nonzero elements of the vector state. An important advantage is that the
Hashmap size is independent of the number of qubits, \textit{e.g}. the state
$\ket5$ can represent the states $\nket{101}$, $\nket{0101}$, and
$\nket{0\dots00101}$ by simply changing the number of qubits, with no
additional memory required. This is at the root of the performance gains
obtained with the bitwise representation. The exponentially growing
system size with the number of qubits does not imply an exponential increase
in memory used to store the \texttt{key}. The Hashmap memory needed grows
only as more basis states are required to describe the system. As we will see
below, the time necessary to operate on the system will also depend on the
Hashmap size, not on the number of qubits. This can generate significant
speedups.

The evolution of a pure state is given by a unitary gate $U$. We represent these with the Hashmap \texttt{bwGate}, which takes an unsigned integer as \texttt{key} and has a set of pairs as the mapped value. Each pair in this set is given by a complex number and an unsigned integer. The \texttt{key} represents a base vector the gate is acting on and the set of pairs represents each output base state and it's associate amplitude, according to
\begin{equation}
\label{eq:bwGate}
  U\ket{i} \equiv \sum_{(\alpha_j, j) \,\in\, \texttt{bwGate}[i]} \alpha_j\ket{j}.
\end{equation}

For an arbitrary gate $U$ prepared as \texttt{bwGate}, Algorithm \ref{alg:bwevol} describes it's application to a \texttt{bwQubits} state. The algorithm iterates over the basis states present in \texttt{bwQubits}' superposition, applying the gate to each of them and constructing the final state.

\subsubsection*{Algorithm 1: arbitrary gate.}
Let \texttt{size} be the number of qubits in a given system we want to evolve, and \texttt{bwGateSize} the number of qubits this gate acts on. The current version of \qs\, only allows user-defined multi-qubit gates applied to neighboring qubits, where \texttt{qubitID} labels the first affected qubit. We remark that more general gates would have no additional computational complexity and are planned for future versions. We now look into each step of the pseudocode in Algorithm \ref{alg:bwevol}. The algorithm starts creating an empty instance of the bitwise Hashmap representation (line \ref{line:11}) and iterates over all values of \texttt{bwQubits}' \texttt{key} (line \ref{line:12}). The \texttt{key} is split into three parts: $x$, $y$ and $z$. Variables $x$ and $z$ hold the qubit values to the left and right, respectively, of the qubits affected by the gate (lines \ref{line:13} and \ref{line:16}). Variable $y$ holds the value of the qubits the gate will operate on. As each gate is defined to operate on a specific system dimension, $y$ must hold the state information on the far-right bits (lines \ref{line:14} and \ref{line:15}).

With the split \texttt{key}, we iterate over the \texttt{bwGate}[$y$] (line
\ref{line:17}), applying $U\ket{y}$. Variables $x$, $i$ (shifted back to
\texttt{qubitID}), and $z$ are concatenated with a bitwise OR (line
\ref{line:18}), with variable \texttt{xiz} representing one output key of
$U\ket{y}$. The output state is updated according to Eq.(\ref{eq:bwGate})
(line \ref{line:19}) and any key for which the state has zero amplitude is
removed from the Hashmap (lines \ref{line:110} and \ref{line:111}). Finally
the output of the gate is returned (line \ref{line:112}).
\begin{algorithm}[h]
  \caption{Arbitrary gate in the bitwise representation.}  
  \label{alg:bwevol}
  \DontPrintSemicolon

  \KwIn{%
  \begin{tabular}[ht]{ll}
      Hashmap($\mathbb{N}\rightarrow\mathbb{C}$)                          & \texttt{bwQubits},   \\
      Hashmap($\mathbb{N} \rightarrow$ Set($\mathbb{C}\times\mathbb{N}$)) & \texttt{bwGate},     \\
      qubitIndex                                                          & \texttt{qubitID}    \\
  \end{tabular}
  }

  \KwOut{\texttt{bwQubits}' quantum state with the \texttt{bwGate} applied at the qubit \texttt{qubitID}.}

  \texttt{newBwQubits} $\leftarrow$ Hashmap($\mathbb{N}\rightarrow\mathbb{C}$)\;                 \label{line:11}

  \ForAll{\normalfont{\texttt{key} $\in$ \texttt{bwQubits}}} {                                   \label{line:12}
    $x  \leftarrow \texttt{key} \wedge ((1 \ll \texttt{qubitID})-1) \ll (\texttt{size-qubitID}))$\; \label{line:13}
    $y' \leftarrow \texttt{key} \gg (\texttt{size-qubitID-bwGateSize})$ \;                        \label{line:14}
    $y  \leftarrow y' \wedge ((1 \ll \texttt{bwGateSize}) -1) $\;                                        \label{line:15}
    $z  \leftarrow \texttt{key} \wedge ((1 \ll (\texttt{size-qubitID-bwGateSize}))-1) $\;                  \label{line:16}

    \ForAll{$\alpha_i, i \in$ \normalfont{\texttt{bwGate}[$y$]}}{                                  \label{line:17}
      \texttt{xiz} = $x|(i \ll (\texttt{size-qubitID-bwGateSize}))|z$\;                           \label{line:18}
      \texttt{newBwQubits}[\texttt{xiz}] += $\alpha_i \cdot\texttt{bwQubits}[\texttt{key}]$\;      \label{line:19}
      \If{\normalfont{\texttt{newBwQubits}[\texttt{xiz}] = 0}}{                                  \label{line:110}
        remove \texttt{xiz} from \texttt{newBwQubits}\;                                          \label{line:111}
      }
    }
  }
  \Return{\normalfont{\texttt{newBwQubits}}}\;                                                   \label{line:112}
\end{algorithm}

The complexity of Algorithm \ref{alg:bwevol} is $O(s\cdot g)$, where $s$ is
the number of computational basis states needed to describe the quantum
state, and $g$ is the number of basis states needed to describe $U\ket{i}$.
This algorithm can be optimized for specific quantum gates, as shown in
Algorithm \ref{alg:h} (Hadamard gate) and Algorithm \ref{alg:cnot} (CNOT
gate). The advantage of these optimizations is that the corresponding gate
matrix does not need to be stored in memory. Note that in these optimizations
$g$ is a constant, so the time complexity is $O(s)$.

\subsubsection*{Algorithm 2: Hadamard gate.}
We now look into an optimized algorithm to apply a Hadamard gate to qubit \texttt{qubitID} in the \texttt{bwQubits} state. The algorithm starts creating an empty instance of the bitwise Hashmap representation (line \ref{line:21}). Then, iterating over  \texttt{bwQubits}' \texttt{key} (line \ref{line:22}), it changes the output Hashmap depending on the value \texttt{qubitID} holds in the \texttt{key}. If this value is 1 (line \ref{line:24}), or 0 (line \ref{line:25}), the complex value of the output state, associated with the same key is updated accordingly.

An auxiliary \texttt{key}' is then created, differing from the original \texttt{key} only by the bit representing \texttt{qubitID}, which is flipped (line \ref{line:28}). The value associated with this new \texttt{key} in the output state is updated (line \ref{line:29}). Any \texttt{key} with zero amplitude is removed from the output Hashmap (lines \ref{line:26}, \ref{line:27}, \ref{line:210} and \ref{line:211}), which is finally returned (line \ref{line:212}).
\begin{algorithm}[h]
  \caption{Hadamard gate in the bitwise representation.}  
  \label{alg:h}

  \DontPrintSemicolon
  \KwIn{%
  \begin{tabular}[t]{ll}
      Hashmap($\mathbb{N}\rightarrow\mathbb{C}$) & \texttt{bwQubits}, \\
      qubitIndex                                 & \texttt{qubitID}    \\
  \end{tabular}
  }

  \KwOut{\texttt{bwQubits}' quantum state with a Hadamard gate applied at the
  qubit \texttt{qubitID}.}

  \texttt{newBwQubits} $\leftarrow$ Hashmap($\mathbb{N}\rightarrow\mathbb{C}$)\;                        \label{line:21}
  \ForAll{\normalfont{\texttt{key} $\in$ \texttt{bwQubits}}}{                                           \label{line:22}
    \eIf{\normalfont{$\texttt{key} \wedge (1 \ll (\texttt{size-qubitID-1}))$}}{                                \label{line:23}
      \texttt{newBwQubits}[\texttt{key}] $-$= $\frac{\textrm{\texttt{bwQubits}[\texttt{key}]}}{\sqrt{2}}$\; \label{line:24}
    }{
      \texttt{newBwQubits}[\texttt{key}] += $\frac{\textrm{\texttt{bwQubits}[\texttt{key}]}}{\sqrt{2}}$\; \label{line:25}
    }
    \If{\normalfont{\texttt{newBwQubits}[\texttt{key}] = 0}}{                                           \label{line:26}
      remove \texttt{key} from \texttt{newBwQubits}\;                                                   \label{line:27}
    }
    \texttt{key}' $\leftarrow$ \texttt{key} $\oplus$ $(1 \ll (\texttt{size-qubitID-1}))$\;                     \label{line:28}
    \texttt{newBwQubits}[\texttt{key}'] += $\frac{\textrm{\texttt{bwQubits}[\texttt{key}]}}{\sqrt{2}}$\;  \label{line:29}
    \If{\normalfont{\texttt{newBwQubits}[\texttt{key}'] = 0}}{                                          \label{line:210}
      remove \texttt{key}' from \texttt{newBwQubits}\;                                                  \label{line:211}
    }
  }
  \Return{\normalfont{\texttt{newBwQubits}}}\;                                                          \label{line:212}
\end{algorithm}

\subsubsection*{CNOT gate in the bitwise}
The bitwise representation allows us to define a very simple algorithm to apply a CNOT gate to the state \texttt{bwQubits}, for a given \texttt{control} and \texttt{target} qubits. Again, the algorithm starts creating an empty instance of the bitwise Hashmap representation (line \ref{line:31}). Then, iterating over \texttt{bwQubits}' \texttt{key} (line \ref{line:32}), if the \texttt{control} bit of the key is 1 (line \ref{line:33}), an auxiliary \texttt{key}' is created, differing from the original \texttt{key} only by the bit representing the \texttt{target} qubit, which is flipped (line \ref{line:34}). This is then used to map the original Hashmap (line \ref{line:35}). If the \texttt{control} qubit is 0, the original Hashmap is copied to the output Hashmap for the same \texttt{key} (line \ref{line:36}). Finally the output Hashmap is returned (line \ref{line:37}).
\begin{algorithm}[bh]
  \caption{CNOT gate in the bitwise representation.}  
  \label{alg:cnot}

  \DontPrintSemicolon
  \KwIn{%
  \begin{tabular}[t]{ll}
      Hashmap($\mathbb{N}\rightarrow\mathbb{C}$) & \texttt{bwQubits}, \\
      qubitIndex                                 & \texttt{control},  \\
      qubitIndex                                 & \texttt{target}    \\
  \end{tabular}
  }

  \KwOut{\texttt{bwQubits}' quantum state with a CNOT applied.}

  \texttt{newBwQubits} $\leftarrow$ Hashmap($\mathbb{N}\rightarrow\mathbb{C}$)\;         \label{line:31}
  \ForAll{\normalfont{\texttt{key} $\in$ \texttt{bwQubits}}}{                            \label{line:32}
    \eIf{\normalfont{$\texttt{key} \wedge (1 \ll (\texttt{size-control-1})) \neq 0$}}{         \label{line:33}
      \texttt{key}' $\leftarrow$  \texttt{key} $\oplus$ $(1 \ll (\texttt{size-control-1}))$\;     \label{line:34}
      \texttt{newBwQubits}[\texttt{key}'] $\leftarrow$ \texttt{bwQubits}[\texttt{key}]\; \label{line:35}
    }{
      \texttt{newBwQubits}[\texttt{key}] $\leftarrow$ \texttt{bwQubits}[\texttt{key}]\;  \label{line:36}
    }
  }
  \Return{\normalfont{\texttt{newBwQubits}}}\;                                           \label{line:37}
\end{algorithm}

\subsubsection*{Measurement in the bitwise}
Qubit measurements in the computational base is done with Algorithm \ref{alg:mea}. The algorithm's idea is to iterate over the Hashmap summing the probability $p_0$, of measuring zero. The measurement outcome is then randomly selected weighted by $p_0$ and a new Hashmap is created with the corresponding collapsed state.

The algorithm starts assigning $0$ probability to $p_0$ (line \ref{line:41}).
All \texttt{bwQuibits}' \texttt{}{key}s are then iterated (line
\ref{line:42}) adding the squared module of the Hashmap value to $p_0$, if
the bit correspondent to \texttt{qubitID} is 0 (line \ref{line:43} and
\ref{line:44}). This is the total probability of measuring zero in qubit
\texttt{qubitID}. With this probability, a uniform distribution is used to
randomly chose the measurement outcome (lines \ref{line:45}-\ref{line:49}).

A new, empty Hashmap is initialized (line \ref{line:410}) and filled with all (normalized) \texttt{bwQubits}' \texttt{key} (line \ref{line:413}) matching the measurement outcome (line \ref{line:412}). Finally, the collapsed state and  measurement result are returned (line \ref{line:414}).
\begin{algorithm}[bh]
  \caption{Measurement in the bitwise representation.}
  \label{alg:mea}
  \DontPrintSemicolon

  \KwIn{%
  \begin{tabular}[ht]{ll}
      Hashmap($\mathbb{N}\rightarrow\mathbb{C}$) & \texttt{bwQubits}, \\
      qubitIndex                                 & \texttt{qubitID}  \\
  \end{tabular}
  }
  \KwOut{The collapsed state and the measurement result of the qubit \texttt{qubitID}.}

  $p_0 \leftarrow 0$\;                                                                                    \label{line:41}
  \ForAll{\normalfont{\texttt{key} $\in$ \texttt{bwQubits}}}{                                             \label{line:42}
    \If{\normalfont{\texttt{key} $\wedge$ $(1 \ll (\texttt{size-qubitID-1}))  = 0$}} {                          \label{line:43}
      $p_0$ += abs$(\texttt{bwQubits}[\texttt{key}])^2$\;                                                 \label{line:44}
    }
  }
  \eIf{$p \neq 0$ and \normalfont{random()} $\leq p_0$} {                                                 \label{line:45}
    \texttt{measurement} $\leftarrow 0$\;                                                                 \label{line:46}
    $p \leftarrow p_0$\;                                                                                  \label{line:47}
  }{
    \texttt{measurement} $\leftarrow 1$\;                                                                 \label{line:48}
    $p \leftarrow 1-p_0$\;                                                                                \label{line:49}
  }
  \texttt{newBwQubits} $\leftarrow$ Hashmap($\mathbb{N} \rightarrow \mathbb{C}$)\;                        \label{line:410}
  \ForAll{\normalfont{\texttt{key} $\in$ \texttt{bwQubits}}}{                                             \label{line:411}
    \If{\normalfont{\texttt{key} $\wedge$ $(1 \ll (\texttt{size-qubitID-1}))$ = \texttt{measurement}}}{          \label{line:412}
      \texttt{newBwQubits}[\texttt{key}] $\leftarrow$ $\frac{\texttt{bwQubits}[\texttt{key}]}{\sqrt{p}}$  \label{line:413}
    }
  }
  \Return{\normalfont{\texttt{newBwQubits}, \texttt{measurement}}}                                        \label{line:414}
\end{algorithm}

\section{Benchmark}
\label{sec:benchmark}

To benchmark the execution time performance of \qs\, and specially its
bitwise representation we compared it with three popular quantum circuit
simulator, Qiskit~\cite{qiskit} developed by IBM, Forest
SDK~\cite{smith_practical_2016} from Rigetti Computing and
Cirq~\cite{Gidney2018} developed by Google. In Qiskit we have used both the
qasm\_simulator and the statevector\_simulator (referred to as qasm and statevector from now on). From Forest SDK we used QVM,
which works as a server and uses PyQuil, also from the Forest SDK, as a
client. To run the QVM server we used the rigetti/QVM docker imagen.
\Tref{tab:setup} shows the configuration of the computer used on the
benchmark, and \Tref{tab:version} shows the software versions used.

\begin{table}[h]
\caption{\label{tab:setup} Computer setup used in the benchmarks.}
\centering
\begin{tabular}{@{}ll}
    \toprule
                       & Model\\
                       \cmidrule{2-2}
    CPU             & Intel\textregistered Core$^{\mathrm{TM}}$ i7-8565U CPU @ 1.80GHz \\
    RAM            & 2x 8GB DDR4 (Speed: 2667 MT/s)      \\
    Linux Kernel & 5.4.6-2-MANJARO                     \\
  \bottomrule
\end{tabular}
\end{table}

\begin{table}[h]
\caption{\label{tab:version} Software versions used in the benchmarks.}
\centering
  \begin{tabular}{llll}
    \toprule
                        & Version  & & Version  \\  
                          \cmidrule{2-2}  \cmidrule{4-4}
    Python                          & 3.8.1     & PyQuil              & 2.15.0 \\
    $\mathcal{Q}$System   & 1.2.0     & Docker             & 19.03.5-ce \\
    GCC                              & 9.2.0     & rigetti/QVM       & 1.15.2 \\
    Qiskit Terra                   & 0.11.0   & Cirq                  & 0.6.0   \\
    Qiskit Aer                      & 0.3.4      \\
  \bottomrule
\end{tabular}
\end{table}

As the bitwise representation promises speedups based on the number of computational basis states required to store the quantum state, we chose four benchmarks well suited to test this feature. These are the state preparation of an $n$-qubit (i) GHZ state, (ii) equal superposition state of all computational basis state and (iii) bipartite entangled superposition $\sum\ket{k}\ket{k}$. Our final benchmark prepares state (ii) and subsequently measures all qubits, returning an $n$-bit string with the measurement outcomes. As these states all differ in their amount of superposition and entanglement, two important resources for quantum computing, they give us a tool to determine scenarios where use of the bitwise representation should be beneficial. We remarks that these tests are inline with other recently used benchmarks for quantum circuit simulators~\cite{rogerluo2020}.

\subsubsection*{GHZ state benchmark.}
\begin{figure}[h]
  \subfloat[\label{fig:ghz}Benchmark for GHZ state preparation.]{\includegraphics[width=.48\linewidth]{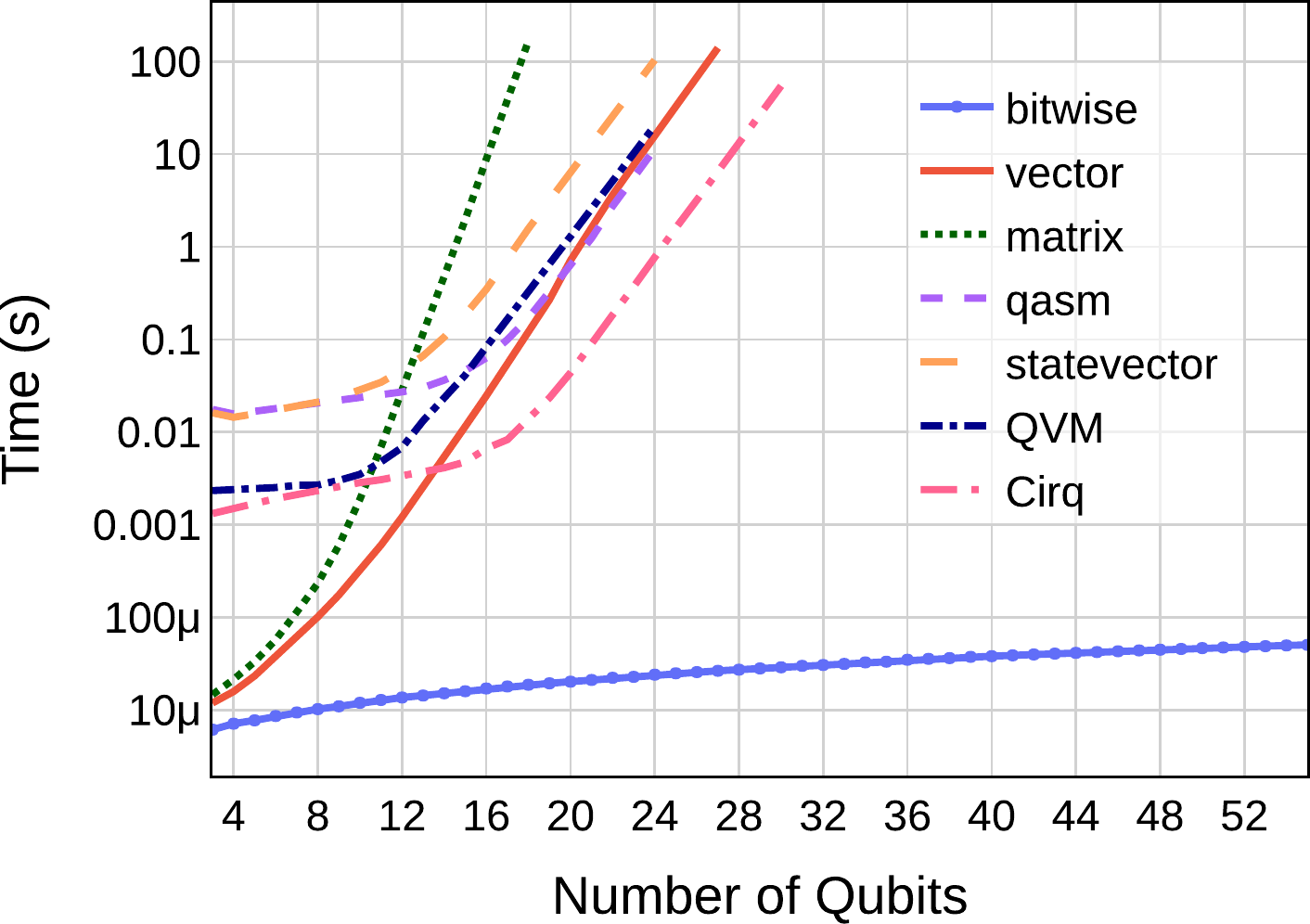}}
  \hfill
  \subfloat[\label{fig:h}Benchmark for the preparation of an equal superposition state, \Eref{eq:eqSuperpos}.]{\includegraphics[width=.48\linewidth]{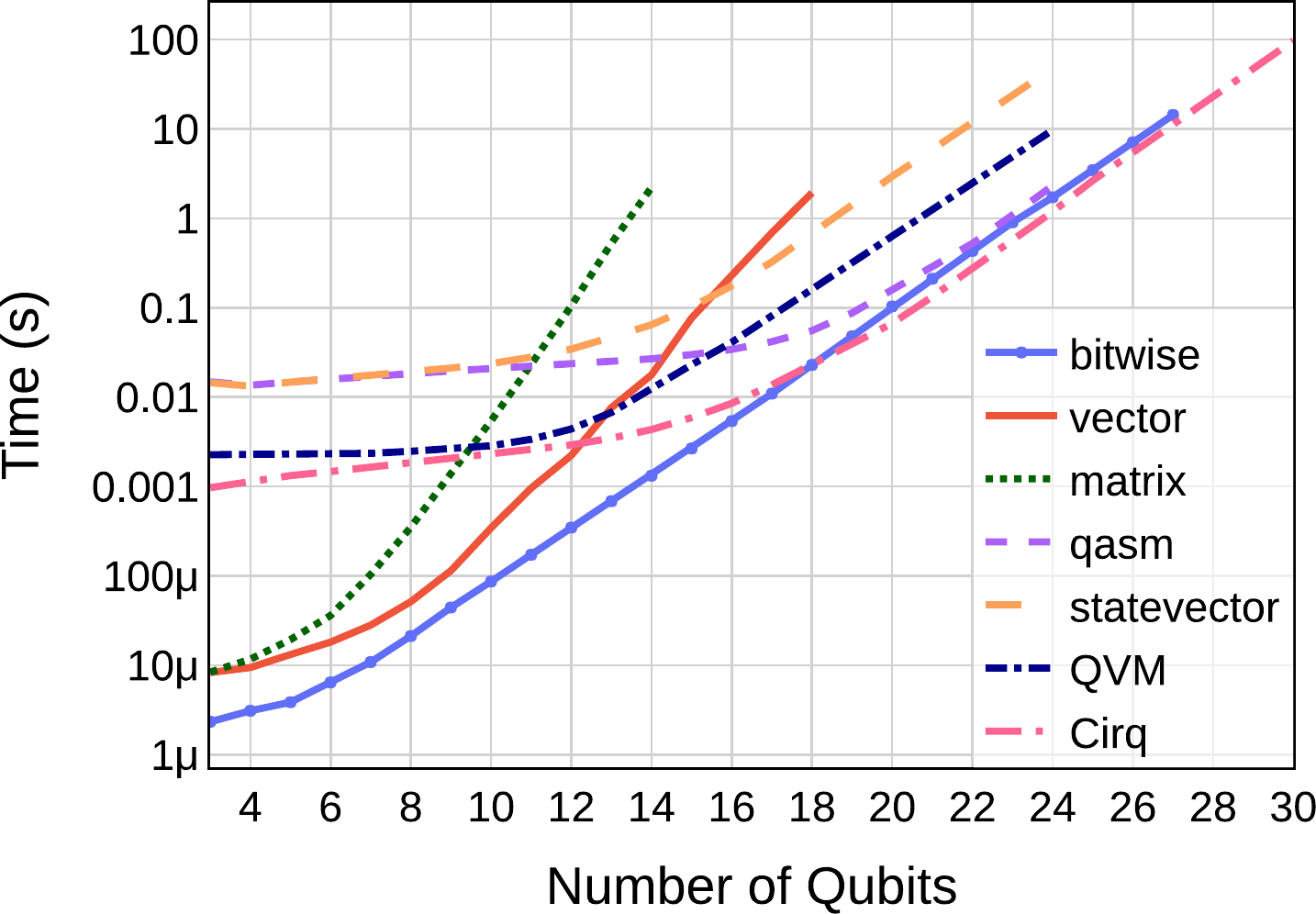}}\caption{}
\end{figure}

The first benchmark prepares an $n$-qubit GHZ state starting from an initial state with all qubits in the $\ket0$ state. The GHZ state has the form
\begin{equation}
\ket{\mathrm{GHZ}}=\frac{1}{\sqrt{2}}\left(\ket{0}^{\otimes n}+\ket{1}^{\otimes n}\right),
\end{equation}
and has a well known circuit for its state preparation. This circuit starts applying a Hadamard gate to a given qubit, which is then used as the control qubit for $(n-1)$ CNOT gates with the other qubits as target. Notice that at any point during the circuit the state is described by only two basis states. The benchmark results are displayed in \Fref{fig:ghz}, which confirms the exponential gain expected for the bitwise representation. As the number of gates grows linearly with the number of qubits (for each new qubit a new CNOT is required), but the size of the Hashmap remains unchanged, we observe a linear scaling of this representation.

\subsubsection*{Equal superposition state benchmark.}
We now move to a benchmark where no gain is expected for the bitwise representation: the preparation of a state with an equal superposition of all computational basis states, given by
\begin{equation}
\ket{\psi}=\frac{1}{\sqrt{2^n}}\sum_{k=0}^{2^n-1}\ket{k}.
\label{eq:eqSuperpos}
\end{equation}
The circuit used here consists of the application of a Hadamard gate to all qubits, which were initialized in their $\ket0$ state. While the number of required gates again scales linearly with the number of qubits, the number of basis states required scales exponentially. As expected, this state preparation suffers from an exponential time scaling, shown in \Fref{fig:h}. However, we note that this scaling is comparable to the Cirq simulator, which had the best performance.

\subsubsection*{Entangled registers benchmark.}
The next benchmark is for the preparation of state
\begin{equation}
\ket{\psi}=\frac{1}{\sqrt{2^n}}\sum_{k=0}^{2^n-1}\ket{k}\otimes\ket{k}.
\label{eq:entRegisters}
\end{equation}
This can be seen as a bipartite state with each partition containing $n$ qubits. The circuit starts with two registers with $n$ qubits each, all initialized to $\ket0$. The circuit described above is then applied to the first register, creating the state 
\begin{equation}
\ket{\psi}=\left(\frac{1}{\sqrt{2^n}}\sum_{k=0}^{2^n-1}\ket{k}\right)\otimes\left(\ket{0}^{\otimes n}\right).
\end{equation}
After that, CNOTs are applied between the $i$-th qubit of each register, having the $i$-th qubit in the first register as control and the $i$-th qubit in the second register as target, thus creating the desired state of \Eref{eq:entRegisters}. This is an interesting benchmark as the desired state has features of both previous benchmarks. The first part of the state preparation is precisely the one used above, with an observed exponential scaling, but the second part does not change the size of the Hashmap holding the quantum state. This happens because each basis state is mapped to a unique new one, \textit{e.g.} $\ket k\otimes\ket{0}^{\otimes n}\rightarrow\ket k\otimes\ket k$.

\begin{figure}[h]
  \subfloat[\label{fig:hcnot}Benchmark for the preparation of entangled registers, \Eref{eq:entRegisters}.]{\includegraphics[width=.48\linewidth]{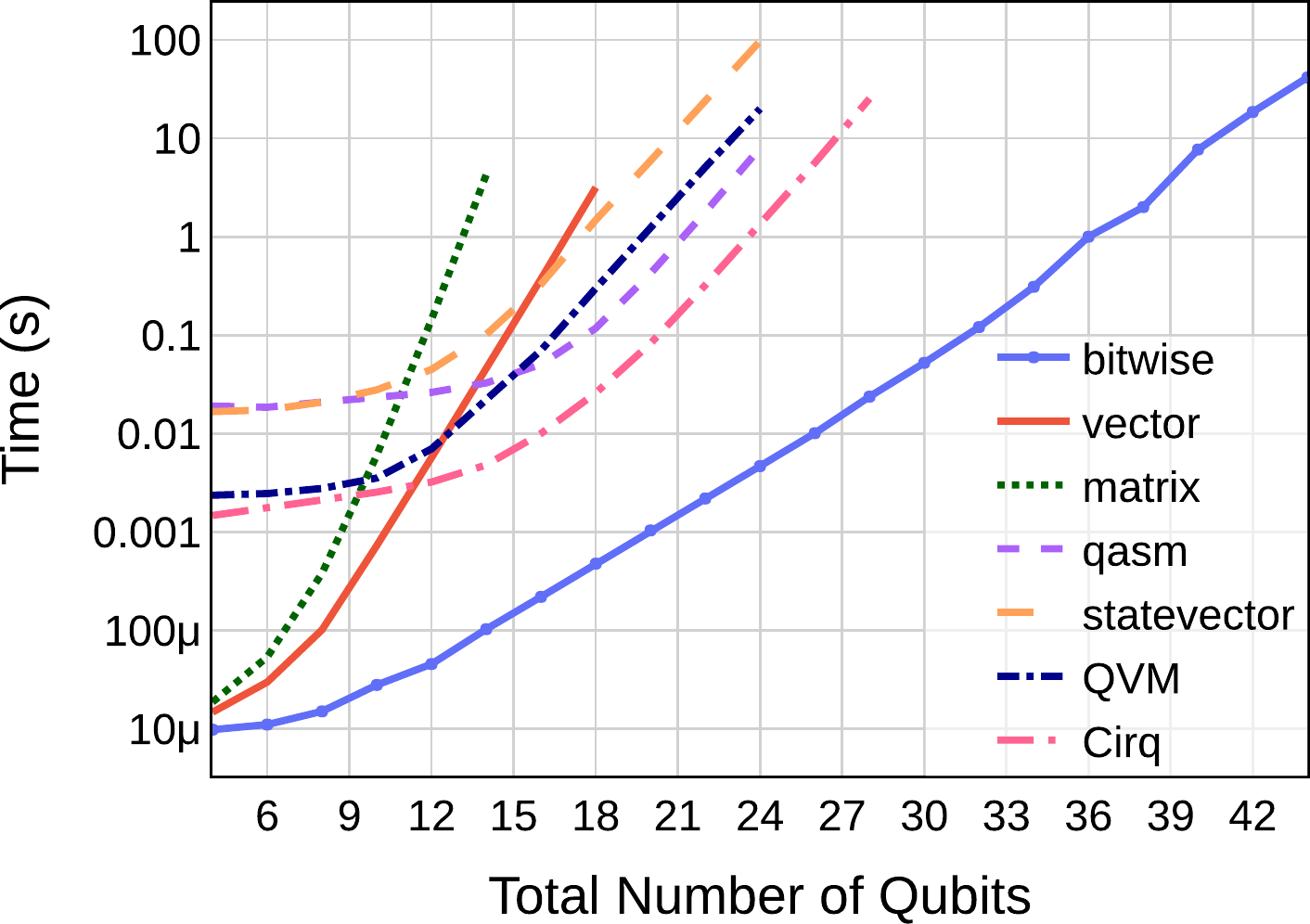}}
  \hfill
  \subfloat[\label{fig:mea}Benchmark for measurement of all qubits for an initial state given by \Eref{eq:eqSuperpos}.]{\includegraphics[width=.48\linewidth]{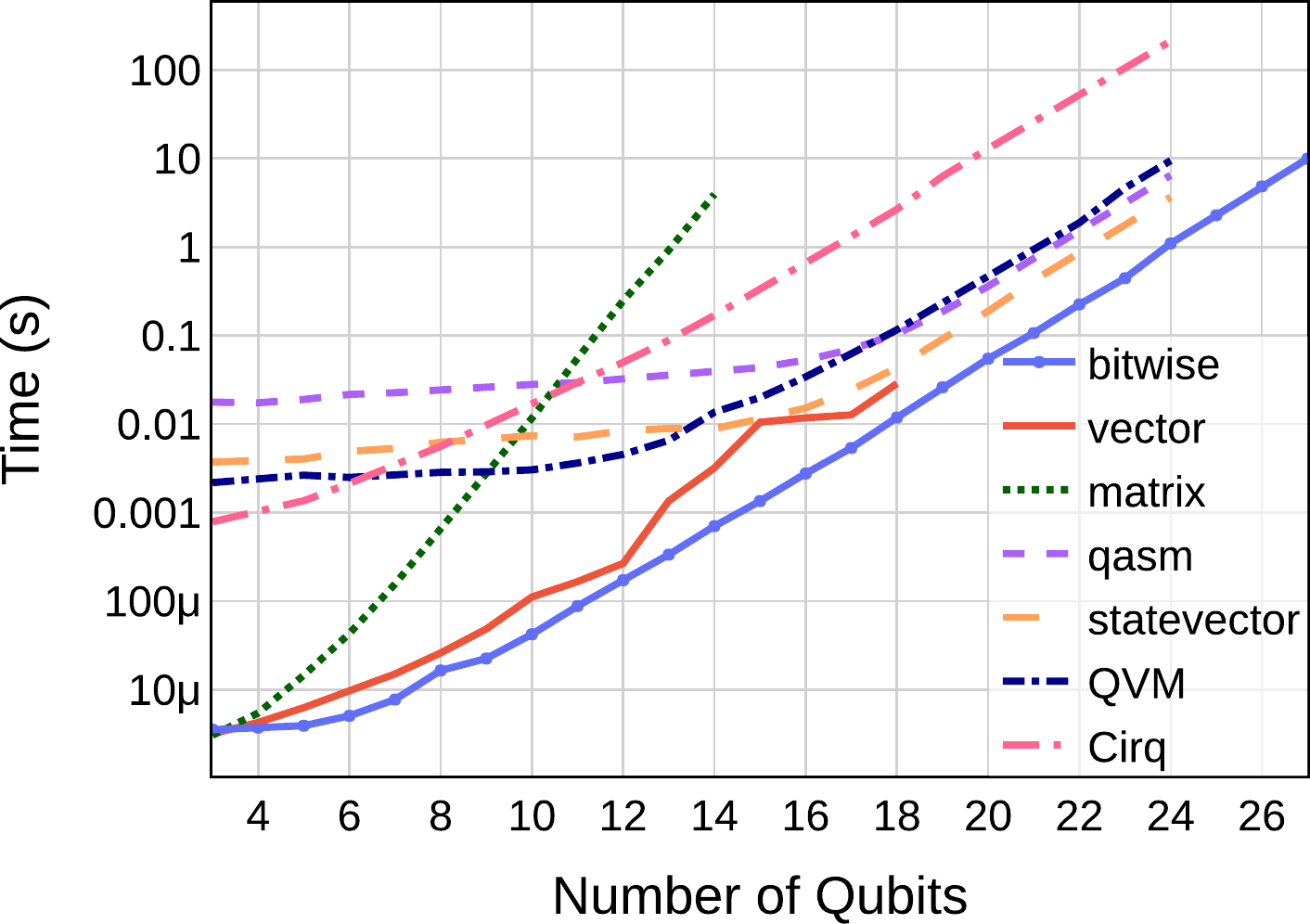}}  \caption{}
\end{figure}

\Fref{fig:hcnot} shows the benchmark results which confirm the exponential scaling of all simulators. However, the bitwise representation displays a much better performance and reduced scaling.

\subsubsection*{Random number measurement}
Our last benchmark is designed to test the efficiency of the measurement scheme in the bitwise representation. The circuit first prepares a test state, in this case the equal superposition state of \Eref{eq:eqSuperpos}, and then measures each qubit, returning the collapsed state, as well an $n$-bit string with the measurement results. In the bitwise representation the measurement scheme is a sequential application of Algorithm~\ref{alg:mea} to each qubit. \Fref{fig:mea} shows the time taken for the measurement circuit, determined by the subtraction of the complete run time and the time used to create the state, as displayed in \Fref{fig:h}. With the exception of the matrix simulators, all other simulators show similar exponential scalings. However, the bitwise operations show an overhead reduction of more than an order of magnitude when compared to Cirq. We note that an optimization flag used in QVM was turned off in order to clearly determine the measurement time. The measurement algorithm was not optimized to take into account the multiple simultaneous measurements. This could done by, \textit{e.g.}, determining the probabilities of all measurement outcomes in a single loop through the Hashmaps \texttt{key}s and selecting the obtained results form this distribution.

\section{Conclusion}
\label{sec:conclusion}
We presented the bitwise representation as a powerful tool for quantum circuit simulations. This was implemented in the more general framework of $\mathcal{Q}$System, a quantum circuit simulator for Python that can be used to study and develop quantum algorithms, protocols and error studies. The bitwise representation takes advantage of a Hashmap to store quantum states and gates and bitwise logical and shift operations to speedup simulations. Benchmarking against popular simulators such as Qiskit, Forest SDK and Cirq, we showed that, as predicted from its construction model, this operations show exponential speedups when the simulated state does not require a large number of computational basis states to be fully described. It is also interesting to notice that even in situations where no speedup was expected, an exponential scaling comparable to the other simulators was obtained.

It is important to notice that the exponential gain does not depend on any knowledge of which basis states are required, but only on the fact that the size of this set is not exponentially increasing. Moreover, while the current implementation on \qs\, uses the Pauli Z operator to define the computational basis states, other basis states can be considered for problems known to be more efficiently described in other basis sets. One such possibility could be to implement similar bitwise implementations to simulate circuits with states well described by Matrix Product States~\cite{PerezGarcia:2006}. It is also interesting to see that entanglement is not directly related to the observed speedups, as the biggest gains where obtained for the generation of the genuinely, maximally entangled GHZ state.

While the current implementation relies on Hashmaps to hold the quantum state, the bitwise representation can also be implemented on other data structure types. This may prove helpful as Hashmaps have a high memory usage. The current version of \qs\, was not optimized for memory use.

Detailed documentation for \qs\, can be found in Reference~\cite{Rosa2019} and the full numerical simulations are available in Reference~\cite{Rosa2020}.

\ack
The authors are grateful to J. Marchi for helpful discussions and T. O. Maciel for constructive criticism of the manuscript. B.G.T. acknowledges support from FAPESC and CNPq INCT-IQ (465469/2014-0). E.C.R.R. acknowledges support from CAPES - Finance Code 001.

\section*{References} 
\bibliography{main}

\end{document}